\documentclass[11pt,a4paper]{article}
\usepackage[hyperref]{emnlp2020}
\usepackage{times}
\usepackage{latexsym}

\usepackage{microtype}

\usepackage{amssymb}
\usepackage{graphicx}
\usepackage{enumitem}
\usepackage{subfig} 
\usepackage{amsmath}
\usepackage{booktabs}

\usepackage{algorithm}
\usepackage{algorithmic}
\usepackage{multirow} 
\usepackage{amsmath} 
\usepackage{xcolor} 

\aclfinalcopy 

\title{Named Entity Recognition Only from Word Embeddings}

\author{Ying Luo, Hai Zhao \thanks{$\ $ Corresponding author. This paper was partially supported by National Key Research and Development Program of China (No. 2017YFB0304100), Key Projects of National Natural Science Foundation of China (U1836222 and 61733011), Huawei-SJTU long term AI project, Cutting-edge Machine reading comprehension and language model.} \and Junlang Zhan\\
Department of Computer Science and Engineering, Shanghai Jiao Tong University \\
Key Laboratory of Shanghai Education Commission for Intelligent Interaction \\ and Cognitive Engineering, Shanghai Jiao Tong University, Shanghai, China\\
MoE Key Lab of Artificial Intelligence, AI Institute, \\ Shanghai Jiao Tong University, Shanghai, China\\
{\tt kingln@sjtu.edu.cn, zhaohai@cs.sjtu.edu.cn}, \\ {\tt longmr.zhan@sjtu.edu.cn}
}

\date{}

\begin{document}
\maketitle
\begin{abstract}
Deep neural network models have helped named entity recognition achieve amazing performance without handcrafting features. 
However, existing systems require large amounts of human annotated training data. 
Efforts have been made to replace human annotations with external knowledge (e.g., NE dictionary, part-of-speech tags),  while it is another challenge to obtain such effective resources.  
In this work, we propose a fully unsupervised NE recognition model which only needs to take informative clues from pre-trained word embeddings.
We first apply Gaussian Hidden Markov Model and Deep Autoencoding Gaussian Mixture Model on word embeddings for entity span detection and type prediction, and then further design an instance selector  based  on  reinforcement  learning to distinguish positive sentences from noisy sentences and then refine these coarse-grained annotations through neural networks.
Extensive experiments on two  CoNLL benchmark NER datasets (CoNLL-2003 English dataset and CoNLL-2002 Spanish dataset) demonstrate that our proposed light NE recognition model achieves remarkable performance without using   any annotated lexicon or corpus.
\end{abstract}

\section{Introduction}
Named Entity (NE) recognition is a major natural language processing task that  intends to identify words or phrases that contain the names of PER (Person), ORG (Organization),
LOC (Location), etc. 
Recent advances in deep neural models allow us to build reliable NE recognition systems \cite{Lample:16,Ma:16,liu2018empower,yang2018ncrf,luo2019hierarchical,luo-zhao-2020-bipartite}.
However, these existing methods require large amounts of manually annotated data for training supervised models.    
There have been efforts to deal with the lack of annotation data in NE recognition, 
\cite{talukdar2010experiments}  train a weak supervision model and use label propagation methods to identify more entities of each type;
\cite{shen2017deep} employ Deep Active Learning to efficiently select the set of samples for labeling, thus greatly reduce the annotation budget; 
\cite{ren2015clustype,Shang:18,fries2017swellshark,yang2018distantly,Jie2019} use partially annotated data or external resources such as NE dictionary, knowledge base, POS tags as a replacement of hand-labeled data to train  distant supervision systems.
However, these methods still have certain requirements for annotation resources. 
Unsupervised models have achieved excellent results in the fields of part-of-speech induction \cite{lin2015unsupervised,stratos2016unsupervised}, dependency parsing \cite{he2018unsupervised,pate2016grammar}, etc.  
Whereas the development of unsupervised NE recognition is still kept unsatisfactory.
\cite{Liu2019} design a Knowledge-Augmented Language Model for unsupervised NE recognition, they perform NE recognition by controlling whether a particular word is modeled as a general word or as a reference to an entity in the training of language models. However, their model still requires type-specific entity vocabularies for computing the type probabilities and the probability of the word under given type.

Early unsupervised NE systems relied on labeled seeds and discrete features \cite{collins1999unsupervised}, open web text \cite{etzioni2005unsupervised,nadeau2006unsupervised}, shallow syntactic knowledge \cite{zhang2013unsupervised}, etc.
Word embeddings learned from unlabeled text provide representation with rich syntax and semantics and have  shown to be valuable as features in unsupervised learning problems \cite{lin2015unsupervised,he2018unsupervised}. 
In this work, we propose an NE recognition model with word embeddings as the unique feature source. 
We separate the entity span detection and entity type prediction into two steps.
We first use Gaussian Hidden Markov Model (Gaussian-HMM) to learn the latent Markov process among NE labels with the IOB tagging scheme and then feed the candidate entity mentions to a Deep Autoencoding Gaussian Mixture Model (DAGMM) \cite{zong2018deep} for their entity types.  We further apply BiLSTM and an instance selector based on reinforcement learning \cite{yang2018distantly,feng2018reinforcement} to refine annotated data.
Different from existing distant supervision systems \cite{ren2015clustype,fries2017swellshark,Shang:18,feng2018reinforcement}, which generate labeled data from NE lexicons or knowledge base which are still from human annotation, our model may be further enhanced by automatically labeling data with Gaussian-HMM and DAGMM.

The contribution of this paper is that we propose a fully unsupervised NE recognition model that depends on no external resources or annotation data other than word embeddings. 
The empirical results show that our model achieves remarkable results on CoNLL-2003 English and CoNLL-2002 Spanish benchmark datasets.

The rest of this paper is organized as follows.
The next section introduces our proposed basic
model in detail. Section 3 further gives a refinement model. Experimental results are reported in
Section 4, followed by related work in Section 5.
The last section concludes this paper.

\begin{figure*}[!t]
  \centering 
  \includegraphics[width=.998\linewidth]{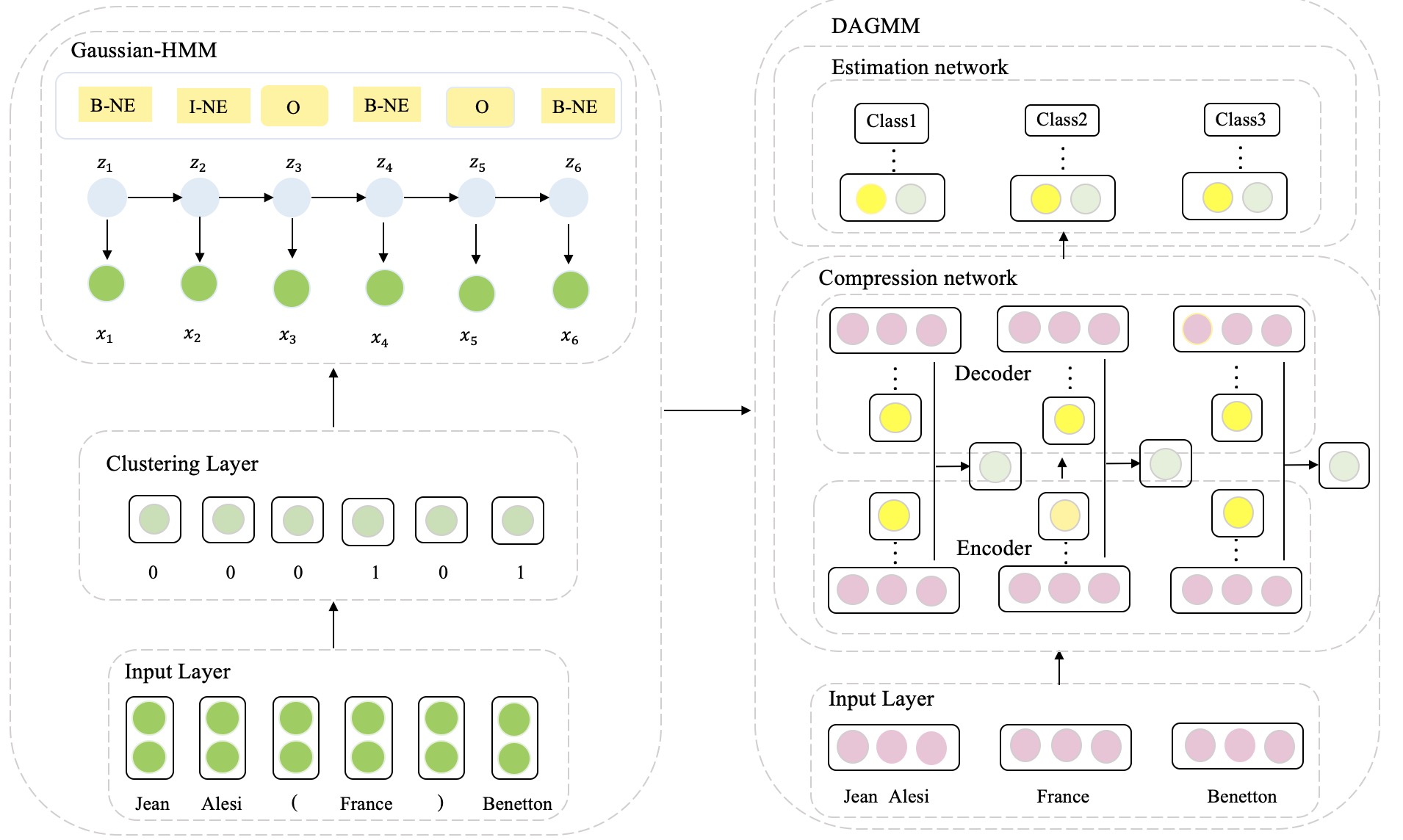}
    \caption{Architecture of the unsupervised NE recognition model. The left part is designed for entity span detection and the right part is used for entity type prediction.}\label{framework}
\end{figure*} 

\section{Model} 
As shown in Figure \ref{framework}, 
the first layer of the model is a two-class clustering layer, which initializes all the words in the sentences with 0 and 1 tags, where 0 and 1 represent non-NE and NE, respectively. 
The second layer is a Gaussian-HMM used to generate the boundaries of an entity mention with IOB tagging (Inside, Outside and Beginning). 
The representation of each candidate entity span is further fed into a Deep Autoencoding Gaussian Mixture Model (DAGMM) to identify the entity types.

\subsection{Clustering}
 The objective of training word embeddings is to let words with similar context occupy close spatial positions. 
 \cite{Miran:16} conduct experiments on the nearest neighbors of NEs and discover that similar NEs are more likely to be their neighbors, since NEs are more similar in position in the corpus and syntactically and semantically related.
 Based on the discoveries above, we perform K-Means clustering algorithm on the word embeddings of the whole vocabulary. 
 According to the clusters, we assign words in the cluster with fewer words 1 tags, and the other cluster 0 tags (according to the statics of \cite{Jie2019}, the proportion of NEs is very small on CoNLL datasets), and generate a coarse NE dictionary using the words with 1 tag.

\subsection{Gaussian HMM} 
Hidden Markov model is a classic model for NE recognition \cite{zhou2002named,zhao2004named}, since hidden transition matrix exists in the IOB format of the NE labels \cite{sarkar2015hidden}. 
We follow the Gaussian hidden Markov model introduced by \cite{lin2015unsupervised,he2018unsupervised}. 
Given a sentence of length $l$, we denote the latent NE labels as $ z = \{z_i\}_{i=1}^l$, the cluster embeddings as $v = \{v_i\}_{i=1}^l $,
observed (pre-trained) word embeddings as $x = \{x_i\}_{i=1}^l$, transition parameters
as $\theta$. The joint distribution of observations and latent labels is given  as following:
\begin{equation}  
    p(z, x, v; \theta) = 
    \prod_{i=1}^l p(z_i | z_{i-1}; \theta) p(x_i | z_i)p(v_i| z_i)
\end{equation}  
where $p(z_i | z_{i-1}; \theta) $ is the multinomial transition probability, $p(x_i | z_i)$ is the multivariate emission probability, which represents the 
probability of a particular label generating the embedding at position $i$. 

Cluster features (0, 1 tags) carry much word-level categorization information and can indicate the distribution representation, which we map to 3-dimension cluster embeddings $v \in \mathbb{R} ^{2 \times 3}$. 
We initialize $v^{2 \times 3}$ as $[[1, 0, 0], [0, 0.5, 0.5]]$ (corresponding to \emph{O}, \emph{I}, \emph{B} tag, respectively), which means that if the cluster tag of a word is 0, we initialize the word with all the probability of being \emph{O} tag, otherwise it will be half of the probability of being \emph{B} or \emph{I} tag. $p(v_i| z_i)$ is obtained through this lookup table, and we fine-tune the cluster embeddings during the training. 

. 

\noindent \textbf{Gaussian emissions} 
Given a label $z \in \{B, I, O\}$, we adopt multivariate Gaussian distribution with 
 mean $\mu_z$ and covariance matrix 
$\Sigma_z$ as the emission probability. The conditional probability density is in a form as:
\begin{equation}  
p(x; \mu_z, \Sigma_z) = \frac{exp(-\frac{1}{2}(x-\mu_z)^T\Sigma_z^{-1}(x - \mu_z))}{\sqrt{(2\pi)^d|\Sigma_z|}}
\end{equation} 
where $d$ is the dimension of the embeddings, $| \cdot |$ denotes the determinant of a matrix. 
The equation assumes that embeddings of words labeled as $z$ are concentrated around the point $\mu_z$, and the concentration is attenuated according to the covariance matrix $\Sigma_z$.

The joint distribution over a sequence of 
observations $x$, cluster sequence $v$ and the latent label sequence $z$ is:
\begin{equation} 
\begin{split}
    p(z, x, v; \theta, \mu_t, \Sigma_t) = \\ \prod_{i=1}^l p(z_i | z_{i-1}; \theta) p(x; \mu_z, \Sigma_z)p(v_i | z_i)
\end{split}
\end{equation}
We use forward algorithm to calculate the probability of $x$ which we maximize during training.

We present two techniques to refine the output of Gaussian-HMM.

 \textbf{Single-word NEs} 
We check the experimental results of Gaussian-HMM and discover that they perform well on the recognition of multi-word NEs, but inferiorly on single-word NEs, which incorrectly gives many false-positive labels, so we need to do further word-level discrimination.  
For a  single-word NE identified by the above model, if it is less than half of the probability of being marked as an NE in the corpus and does not appear in the coarse NE dictionary generated in the clustering step, then we modify it to a non-NE type. Through this modification, the precision is greatly improved without significantly reducing the recall. 

\textbf{High-Quality phrases} Another issue of the above models is the false-negative labels, a long NE may be divided into several short NEs, in which case we need to merge them with phrase matching. 
We adopt a filter to determine high quality phrases according to word co-occurrence information in the corpus:
\begin{equation}
\frac{p ({word_{last}}, {word_{current}})}{ p ({word}_{last}) * p({word}_{current}) } * n > T
\end{equation} 
where $p(\cdot)$ represents the frequency of one word appearing in the corpus, $n$ is the total number of words and $T$ is the threshold, which is set as the default value in word2vec\footnote{https://code.google.com/archive/p/word2vec} for training phrase embeddings.  The intuition behind this is that
if the ratio of the co-occurrence frequency of two
adjacent words to their respective frequencies is
greater than the threshold, then we consider that
these two words are likely to form a phrase. Being
aware of these high-quality phrases, we expect to
enhance the recall of our model.

 After obtaining the candidate entity span mentions, we represent them by separating words in them into two parts, the boundary and the internal \cite{sohrab2018deep}.
The boundary part is important to capture the contexts surrounding the region, we directly take the word embedding 
as its representation. For the internal part, we simply average the embedding of each word to treat
them equally. In summary, given the word embeddings $x$, we obtain the representation $u$ of $NE(i, j)$ as follows:
\begin{equation}   
u = NE(i, j) = [x_i; \frac{1}{j-i + 1}\sum_{k=i}^jx_k;x_j] 
\end{equation} 

\begin{figure*}[!t]
  \centering 
  \includegraphics[width=.75\linewidth]{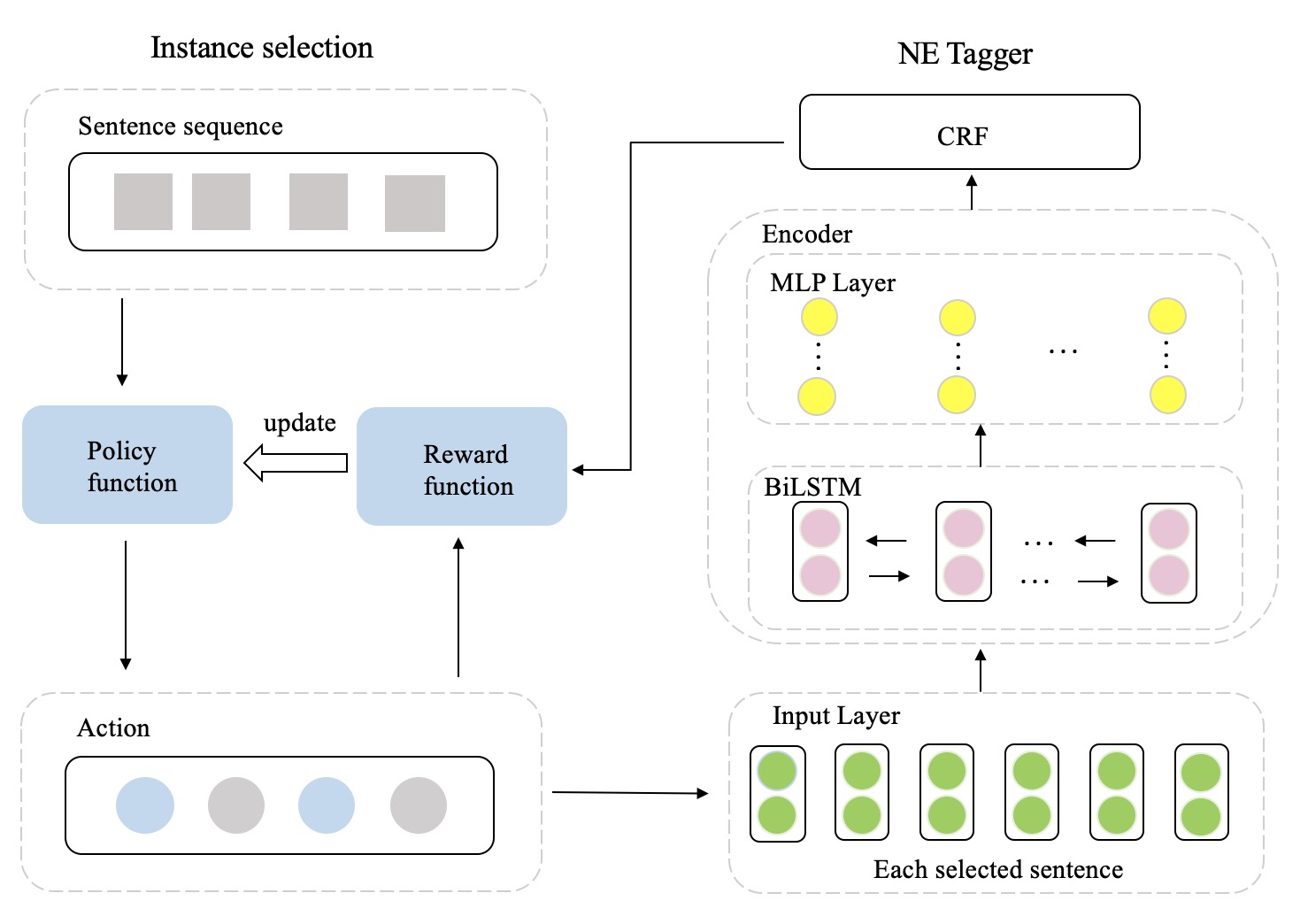}
    \caption{The framework of the reinforcement  learning model, which consists of two parts. The left instance selector filters sentences according to a policy function, and then the selected sentences are used to train a better NE tagger. The instance selector updates its parameters based on the reward computed from NE tagger.  }\label{RL}
\end{figure*} 

\subsection{DAGMM}
 After obtaining the candidate entity mentions, we need to further identify their entity types. Gaussian Mixture Model (GMM) is adopted to learn the distribution of each entity type. Experimental result of \cite{zong2018deep} suggested to us that it is more efficient to perform density estimation in the low-dimensional space, in which case the distribution of words are denser and more suitable for GMM.
 Therefore, we adopt Deep Autoencoding Gaussian Mixture Model (DAGMM) \cite{zong2018deep} to identify NE  types.
 DAGMM consists of two major components: compression network utilizes a deep autoencoder to perform dimension reduction and concatenate the reduced low-dimensional representation and the reconstruction error features as the representations for the estimation network; The estimation network takes the low-dimension representation as input, and uses GMM to perform density estimation.

\textbf{Compression network} contains an encoder function for dimension reduction and a decode function for reconstruction, both of which are multi-layer perceptron (MLP), and we use tanh function as the activation function. Given NE representation $u $, the compression generates its low-dimensional representation $t$ as follows.
\begin{equation}
    \begin{aligned}
        t_e &= MLP(u; \theta_e) \quad  u' = MLP(t_e;\theta_d) \\
        t_r &= f(u, u') \quad\quad\quad t = [t_e, t_r]
    \end{aligned}
\end{equation}
 where $\theta_e$ and $\theta_d$ are respectively the parameters of the encoder and decoder, $u'$ is the reconstruction counterpart of $u$, $f(\cdot)$ denotes the reconstruction error, we take the concatenation of relative Euclidean distance and cosine similarity as $t_r$ in our experiment. $t$ is then fed into the input layer of estimation network. Intuitively, we need to make the reconstruction error low to ensure that the low-dimensional representations preserve the key information of the NE representations. Thus the reconstruction error is taken as part of the loss function and is designed as the $L_2$-norm.
\begin{equation}
    L(u_i, u_i') = \|u_i - u'_i\|_2^2
\end{equation}

\textbf{Estimation network} contains an MLP to predict the mixture
membership for each instance and a GMM with unknown mixture-component distribution $\phi$, mixture means $\mu$ and covariance matrix $\Sigma$ for density prediction.
During the training phase, the estimation network estimates the parameters of GMM and evaluates the likelihood for the instances. Given the low-dimensional representation $t$ and the number of entity types $K$ as the number of mixture components, MLP maps the representation to the $K$-dimension space:
\begin{equation}
    \begin{aligned}
        m = MLP(t;\theta_m) \\
        \hat{\gamma} = softmax(m)
    \end{aligned}
\end{equation}
where $\theta_m$ is the parameter of MLP, $\hat{\gamma}$ is a $K$-dimension vector for the soft mixture-component membership prediction. The  estimation network estimates the parameters of GMM as follows ($\forall 1 \le k \le K$),
\begin{equation} 
    \begin{aligned}
        \hat{\phi}_k = \sum_{i=1}^N\frac{\hat{\gamma}_{ik}}{N}, 
        \hat{\mu}_k = \frac{\sum_{i=1}^N\hat{\gamma}_{ik}t_i}{\sum_{i=1}^N\hat{\gamma}_{ik}} \\
        \hat{\Sigma}_k = \frac{\sum_{i=1}^N\hat{\gamma}_{ik}(t_i - \hat{\mu}_k)(t_i - \hat{\mu}_k)^T}{\sum_{i=1}^N\hat{\gamma}_{ik}}
    \end{aligned}
\end{equation}
where $\hat{\gamma_i}$ is the membership prediction for $t_i$, and $\hat{\phi}_k, \hat{\mu_k}, \hat{\sigma_k}$ are
mixture probability, mean, covariance for component $k$ in GMM, respectively.

The likelihood for the instance is inferred by
\begin{equation}
    E(t) = -log(\sum_{k=1}^K\hat{\phi}_k \frac{exp(-\frac{1}{2}(t-\hat{\mu}_k)^T\hat{\Sigma}_k^{-1}(t-\hat{\mu}_k)}{\sqrt{(2\pi)^d|\hat{\Sigma}_k}|})
\end{equation} 

To avoid the diagonal entries in covariance matrices degenerating to 0, we penalize small values on the diagonal entries by
\begin{equation}
    p(\hat{\Sigma}) = \sum_{k=1}^K \sum_{j=1}^d \frac{1}{\hat{\Sigma}_{kjj}}
\end{equation}
where $d$ is the dimension of the $t$.

During training,  we minimize the joint objective function:
\begin{equation}
    \begin{aligned}
        J(\theta_e, \theta_d, \theta_m) &= \frac{1}{N}\sum_{i=1}^NL(u_i, u'_i) \\ 
        & + \frac{\lambda_1}{N}\sum_{i=1}^NE(t_i) + \lambda_2 P(\hat{\Sigma})
    \end{aligned}
\end{equation}
where $\lambda_1$ and $\lambda_2$ are two user-tunable parameters. 

The final output is the result of $K$ (the number of entity types)  classification. We can only identify whether a word is an NE and whether several NEs are of the same category,
since the entity type names as any other user-defined class/cluster/type names are just a group of pre-defined symbols by subjective naming. 
Therefore, following most work of unsupervised part-of-speech induction such as \cite{lin2015unsupervised}, we use matching to determine the corresponding entity category of each class, just for evaluation.

\section{Refinement}  
The annotations obtained from the above procedure are noisy, we apply Reinforcement  Learning (RL) \cite{feng2018reinforcement,yang2018distantly} to  distinguish  positive  sentences  from  noisy  sentences and  refine  these  coarse-grained  annotations. The RL model has two modules: an NE tagger and an instance selector.
 
\subsection{NE tagger}
 Given the annotations generated by the above model, we take it as the noisy annotated label to train the NE tagger. 
 Following \cite{Lample:16,yang2018design,yang2018ncrf}, we employ bi-directional Long Short-Term Memory network (BiLSTM) for sequence labeling.
 In the input layer, we concatenate the word-level and character-level embedding as the joint word representation.
 We employ BiLSTM as the encoder, the  concatenation of the forward and backward network output features $[\overrightarrow{h_k}, \overleftarrow{h_k}]$  is fed into an MLP, and then feed the output of MLP to a CRF layer. 
 
CRF \cite{Lafferty:01} has been included in most sota models, which captures label dependencies by adding transition scores between adjacent labels.
During the decoding process, the Viterbi algorithm is used to search the label sequence with the highest probability.
 Given a sentence of length $l$, we denote the input sequence $x = \{x_1,...,x_l\}$, where $x_i$ stands for the $i^{th}$ word in sequence $x$.
For $y=\{y_1,...,y_l\}$ being a predicted sequence of labels for $x$. We define its score as
\begin{equation}  
score(x, y) = \sum_{i=0}^{l-1} T_{y_i, y_{i+1}} + \sum_{i=1}^l P_{i, y_i}
\end{equation}
where $T_{y_i, y_{i+1}}$ represents the transmission score  from the $y_i$ to $y_{i+1}$,
$ P_{i, y_i}$  is the score of the $i^{th}$ tag of the $i^{th}$ word from the BiLSTM.

A softmax over all possible tag sequences in the sentences generates a probability for the sequence $y$:
\begin{equation}
p(y|x) = \frac{e^{score(x,y)}}{\sum_{\tilde{y}\in Y}e^{score(x, \tilde{y})}} 
\end{equation}
where $Y$ is the set of all possible tag sequences. During the training, we consider the maximum log-likelihood of the correct NE tag sequence. 
While decoding, we predict the optimal sequence which achieves the maximum score:

\begin{equation}  
y^{*} = \arg\max_{\tilde{y} \in Y} score(x, \tilde{y})
\end{equation}
  
\subsection{Instance Selector}
The instance selection is a reinforcement learning process, where the instance selector acts as the agent and interacts with the environment (sentences) and the NE tagger, as shown in Figure \ref{RL}. Given all the sentences, the agent takes an action to decide which sentence to select according to a policy network at each state, and receives a reward from the NE tagger when a batch of $N$  sentences have been selected.

\textbf{State representation.}  We follow the work of \cite{yang2018distantly} and represent the state $s_j$ as the concatenation of the serialized vector representations from BiLSTM and the label scores from the MLP layer. 

\textbf{Policy network.} The agent makes an action $a_j$ from set of {\{0, 1\}} to indicate whether the instance selector will select the $j^{th}$ sentence.
We adopt a logistic function  as  the policy function:
\begin{equation}
    \begin{aligned}
    A(s_j, a_j) &= a_i \sigma(W*s_j + b)  \\
    &+ (1 - a_j) ( 1 - \sigma ( W*s_j + b ))
    \end{aligned}
\end{equation}
where $W$ and $b$ are the model parameters, and $\sigma(\cdot)$  stands for the logistic function.

\textbf{Reward.} 
The reward function indicates the ability of the NE tagger to predict labels of the selected sentences and only generates a reward  when all the actions of the selected $N$ sentences have been completed,
\begin{equation}
    r = \frac{1}{N} (\sum_{x,y \in \tilde{H}} \log p(y | x))
\end{equation}
where $\tilde{H}$ represents the set of selected $N$ sentences.

\textbf{Training} During the training phase, we optimize the policy network to maximize the reward of the selected sentences.
The parameters are updated as follows,
\begin{equation}
    \Theta = \Theta + \alpha \sum_{j=1}^N r \nabla_{\Theta}\log A(s_j, a_j)
\end{equation}
where $\alpha$ is the learning rate and $\Theta$ is the parameter of the instance selector.

We train the NE tagger and instance selector iteratively. In each round, the instance selector first selects sentences from the training data, and then the positive sentences are used to train the NE tagger, the tagger updates the reward to the selector to optimize the policy function. Different from the work of \cite{yang2018distantly}, we relabel the negative sentences by the NE tagger after each round, and merge them with the positive sentences for the next selection.

\begin{table*}[!t]
\centering
\resizebox{0.998\linewidth}{!}{ 
\begin{tabular}{cccccccc}
\toprule 
\multirow2*{} && \multicolumn{3}{c}{EN} & \multicolumn{3}{c}{SP} \\
\cmidrule{3-8}
& & Pre & Rec & $F_1$  & Pre & Rec & $F_1$\\
\midrule
{\cite{Lample:16}} & LSTM-CRF & 91.0& 90.8& 90.9 & 85.7& 85.8&  85.8   \\ 
\midrule 
\cite{Jie2019} &  IA-Training$^0$ & 89.0 & 90.1 & 89.5&  81.3 & 82.7 &82.0  \\
\midrule
 \multirow2*{\cite{Liu2019}} & Dict $^1$ & - & - & 72 &- & - &-\\
 & Dict +$P(\tau | y)^2$ & - & - & 76 &- & - &-\\
\midrule
 \multirow3*{\cite{Shang:18}}~   & {Dict-Training$^3$  } & 75.18 & 79.71  & 77.38  & 22.11 & 70.89 &  33.71 \\
~   & {Handcraft } & 23.45  & 26.38 & 24.83 &-  & - & - \\
~   & {SENNA } & 7.09  & 7.0 & 7.036  &-  & - & - \\
\midrule\midrule
 \multirow4*{{ Ours}} 
 & {{basic$^4$}} & 62.57 & 56.83 & 60.76 & 45.35 & 53.41 & 49.05 \\ 
~   & {LSTM-CRF } & 73.15 & 60.02  & 65.94 & 49.99 & 56.76 & 53.16\\
~   & {LSTM-CRF + RL$^5$ } & 74.25 &63.51& 68.64  & 50.61 & 58.36 & 54.31 \\
\bottomrule
\end{tabular}}
\caption{Main results of NE recognition on CoNLL 2003 English (EN) and CoNLL 2002 Spanish (SP) datasets.  Superscript annotations: 0: represents incomplete annotations in training data. 1:  type-specific entity vocabularies extracted from   WikiText-2. 2: a prior type information which was pre-computed from entity popularity information. 3: these three represent the lexicon extracted from training data, human annotated lexicon from Wikipedia corpus and SENNA lexicon.  4: Our basic ouput from GMM without refinement. 5: +RL:  add reinforcement learning with instance selector. } 
\label{mainresult}
\end{table*}

\section{Experiments} 
 We conduct experiments \footnote{Code is available at: https://github.com/cslydia/uNER.} on two standard NER datasets: CoNLL 2003 English dataset \cite{sang2003introduction} and CoNLL 2002 Spanish dataset \cite{Erik2002introduction} that consist of news articles.
These datasets contain four entity types: LOC (location), MISC (miscellaneous), ORG (organization), and PER (person). We adopt the  standard data
splitting and use the micro-averaged $F_1$ score as the evaluation metric. 

\subsection{Setup} 
\textbf{Pre-trained Word Embeddings.} For the CoNLL 2003 dataset, we use the pre-trained 50$D$ SENNA embeddings released by \cite{Collobert:11} and 100$D$ GloVe \cite{pennington2014glove} embeddings for clustering and training, respectively. For CoNLL 2002 Spanish dataset, we train 64$D$ GloVe embeddings with the minimum frequency of occurrence as 5, and the window size of 5. \\
\textbf{Parameters and Model Training.} 
For DAGMM, the hidden dimensions for compression network and estimation network are $[75,15]$ and 10, respectively. For NE Tagger, we follow the work of \cite{yang2018ncrf} and use the default experimental settings. We conduct optimization with the stochastic gradient descent, the learning rate is initially set to 0.015 and will shrunk by 5\% after each epoch.  The number of selected sentences at each time is set as 10.
Dropout \cite{Srivastava2014Dropout} of a ratio 0.5 is applied
for embeddings and hidden states.

\subsection{Compared Methods} 
 {Supervised benchmarks} on each dataset are represented to show the gap between supervised and our unsupervised model without any annotation data or external resources.
 LSTM-CRF \cite{Lample:16} is the state-of-the-art supervised NE recognition model.

 \cite{Jie2019} propose an  approach to tackle the incomplete annotation
problem. This work introduces $q$ distribution to model missing labels instead of traditionally uniform distribution for all possible complete label sequences, and uses  $k$-fold cross-validation  for estimating $q$. They report the result of keeping 50\% of all the training data and removing the annotations of the rest entities together with the O labels for non-NEs.

\cite{Liu2019} proposes a Knowledge-Augmented Language Model (KALM), which recognizes NEs during training language models.  Given type-specific entity vocabularies and the general vocabulary, KALM computes the entity probability of the next word according to its context. This work extracts 11,123 vocabularies from WikiText-2 as the knowledge base. WikiText-2 is a standard language modeling dataset and covers 92.80\% of entities in CoNLL 2003 dataset.

\begin{table}[!ht] 
\resizebox{0.98\columnwidth}{!}{
\centering
\begin{tabular}{|p{2.5cm}|p{2cm}|p{2cm}|} 
\hline 
 \textbf{Category}& SENNA & Handcraft  \\
\hline
Location  & 36,697 & 213,318  \\ 
Miscellaneous   & 4,722 & -   \\
Organization & 6,440 & 11,749  \\ 
Person   & 123,283 & 80,050   \\
\hline
Total & 171,142 & 305,117 \\
\hline
\end{tabular}}
\caption{Number of entries for each category in lexicons for \cite{Shang:18} for comparisons with our model, which need no lexicon.}
\label{dictiionary}
\end{table}

 \cite{Shang:18} propose a distant supervision NE recognition model AutoNER using domain-specific dictionaries. This work designs a \emph{Tie or Break} tagging scheme that focuses on the ties between adjacent tokens. Accordingly, AutoNER is designed to distinguish Break from Tie while skipping unknown positions.
The authors report their evaluation results on datasets from a specific domain and their method needs necessary support from an NE lexicon. 
For better comparisons,  we use the lexicon from the training data, the SENNA lexicon presented by \cite{Collobert:11} and our handcraft lexicon
\footnote{This dictionary is mainly based on Wikipedia corpus.}
as the domain-specific dictionary to re-implement their work on CoNLL-2003 English dataset,  the size of each category in each lexicon is shown in Table \ref{dictiionary}. Due to the resource constraints, we only extract the lexicon in training data without labeling a larger dictionary for wider comparisons for CoNLL-2002.
  
  \begin{table}[!t] 
\centering
\resizebox{0.98\columnwidth}{!}{
\begin{tabular}{cccccc} 
\toprule 
 \textbf{}& LOC & MISC & ORG & PRR & overall  \\
\midrule
basic & 0.75 &0.67& 0.64& 0.83 &0.72 \\ 
$P(\tau | y)$ & 0.81 & 0.67 &0.65 &0.88 &0.76 \\
\midrule
Ours & 0.68& 0.45 & 0.60 & 0.83 & 0.69 \\
\bottomrule
\end{tabular}
}
\caption{Comparisons with \cite{Liu2019} on CoNLL-2003 for each entity type. } 
\label{typeresult}
\end{table}

 \begin{table}[t] 
 \resizebox{0.98\columnwidth}{!}{
\centering
\small
\begin{tabular}{p{1cm}cccccc} 
\toprule 
\multirow2*{} & \multicolumn{3}{c}{EN} & \multicolumn{3}{c}{SP} \\
\cmidrule{2-7} 
& Pre & Rec & $F_1$  & Pre & Rec & $F_1$\\
\midrule
{Cluster} & 0.43 & 0.53 & 0.47 & 0.27 & 0.69 & 0.39 \\ 
{HMM} & 0.80 & 0.72 & 0.76 & 0.54 & 0.70 & 0.63 \\
\bottomrule
\end{tabular}}
\caption{Main results for entity span detection. Cluster is the result before sending to Gaussian-HMM,  
HMM is short for Gaussian-HMM.}
\label{spandetection}
\end{table}

\begin{table*}[t!] 
\resizebox{0.998\textwidth}{!}{
\centering
\begin{tabular}{lcccccccccc}
\toprule
\textbf{Intance 1} \\
\midrule
{Instance} & Newcombe & was& quoted& as &saying& in& Sydney& 's & Daily & Telegraph  \\
{Before RL} & B-PER & O & O & O & O & O & B-LOC & O & B-ORG & I-ORG \\
After RL &B-PER & O & O & O & O & O & B-LOC & O & B-ORG & I-ORG \\
\midrule
golden label &B-PER & O & O & O & O & O & B-LOC & O & B-ORG & I-ORG \\
\midrule \midrule
\textbf{Instance 2} \\ 
\midrule
{Instance} & Thursday &'s &overseas &edition  &of &the &People &'s & Daily \\
{Before RL} & O & O & O & O & O & O &O&O&O  \\
{After RL} & O & O & O & O & O & O & B-ORG & I-ORG & I-ORG  \\
\midrule
golden label & O & O & O & O & O & O & B-ORG & I-ORG & I-ORG \\
\bottomrule
\end{tabular}}
\caption{Example of of two instances before and after Reinforcement Learning (RL).
}\label{example}
\end{table*}

\subsection{Results and Comparisons}
We present $F_1$, precision, and recall scores on both datasets in Table \ref{mainresult}. All the models compared in Table \ref{mainresult} besides ours need extra resources to some extent, like partially annotated training data, NE dictionary, etc. While our model achieves comparable results without using any  resources mentioned above. We compare the prediction results for each entity type with \cite{Liu2019} in Table \ref{typeresult}, and it is shown that our model performs well  in LOC, ORG and PER types. These NEs have specific meanings, and more similar in position and length in the corpus, thus their word embeddings can better capture semantic and syntactic regularities, and thus better represent the words, while MISC includes various entity types which may bring significant confusion  on learning type patterns.
While \cite{Liu2019} better regularize the type information from NE dictionaries and re-trained type information. 

Though \cite{Shang:18} achieves better results when using golden NE dictionary for English, they perform poorly on SENNA and our manually annotated dictionary. Specially, when using the gold NE dictionary for training Spanish dataset, the result  is especially unsatisfactory. According to our statistics, over half of the MISC NEs in CoNLL 2002 Spanish training data are labeled as other types in the same dataset, while the ratio is 28\% in CoNLL 2003 English dataset, thus the results differs a lot in the two datasets.
Our models achieve much better performance than those of \cite{Shang:18} by more than doubling their $F_1$ scores in the general NE dictionary (SENNA and human-labeled Wikipedia dictionary). Besides, our unsupervised NE recognition method is shown more general and gives a more stable performance than the distant supervision model in \cite{Shang:18}, which highly relies on the quality of the support dictionary and the domain relevance of the dictionary to the corpus. 

We acknowledge that there still exists a gap between our unsupervised NE recognition model with the sota supervised model \cite{Lample:16,Jie2019}, but the applicability of unsupervised models and the robustness of resource dependence are unreachable by supervised models.

Table \ref{spandetection} lists the results of entity span detection. Our Gaussian-HMM absorbs informative clue from  clustering, and greatly improves the results of entity span detection. For the English dataset, we apply SENNA embedding, which is trained on English Wikipedia and Reuters RCV1 corpus, thus the result of clustering becomes better, leading to a better result of Gaussian-HMM. While for the Spanish dataset, the embedding is trained on Wikipedia corpus only, which has little connection with the CoNLL-2002 datasets, thus the result is slightly lower. Overall, unsupervised modeling based on word embeddings may be more general and robust than dictionary-based and corpus-based modeling.

\subsection{Discussion}
Our model is good at dealing with common NEs, because their word embeddings well represent meanings, thus leading to a better prediction. However, our model is not very satisfactory in dealing with nested NEs. For example, \emph{South Africa} and \emph{Africa} can be taken as NEs respectively, and  \emph{south} is recognized as \emph{O} labels in most of the other cases, thus in this case, our model makes a bias prediction, and only recognizes \emph{Africa}.
Table \ref{example} shows an example of a positive instance and a negative instance before RL and after RL. 
During the training process, the instance selector takes action to select the first instance for training a silver NE Tagger. Then the  second instance is relabeled after one epoch, and merged with the first instance for the next turn. We can discover that the NE Tagger learns the effective features of the ORG type, and can modify the wrong labels in the second instance. 

\textbf{Using Pre-trained Languages Models.}
We have also tried language models such as ELMo and BERT as encoders for Gaussian-HMM, but their sparse characteristics in high-dimensional space are not conducive to Gaussian modeling. Unsupervised models have fewer parameters and simpler training phase, thus  there is no guarantee that the language model will retain its key properties when it is reduced to low dimensions. We further add the pre-trained language model BERT as the additional embeddings for the NE Tagger to refine the output of Gaussian-HMM and DAGMM, which slightly improves our result to 69.99 for CoNLL-2003 English NER and 56.66 for CoNLL-2002 Spanish NER.

\section{Related work} 

Deep neural network models have helped peoples released from handcrafted features in a wide range of NLP tasks \cite{zhang2019open,li-etal-2018-seq2seq,li-etal-2018-unified,li2019dependency,zhou2019head,xiao2019,zhang2019dcmn+,zhang2020SemBERT,zhang2020sg}. LSTM-CRF \cite{Lample:16,Ma:16} is the most state-of-the-art model for NE recognition.
In order to reduce the requirements of training corpus, distant supervised models \cite{Shang:18,yang2018distantly,ren2015clustype,he2017autoentity,fries2017swellshark} have been applied to NE recognition. Recently, \cite{Liu2019} proposed a Knowledge-Augmented Language Model, which trains language models and at the same time compute the probability of the next word being different entity types according to the context given type-specific entity/general vocabularies. Unlike these existing approaches, our study focuses on unsupervised NE recognition learning without any extra resources.

Noisy data is another important factor affecting the neural network models, reinforcement learning has been applied to many tasks, \cite{feng2018reinforcement} use reinforcement learning for Relation Classification from Noisy Data. \cite{yang2018distantly} show how to apply reinforcement learning in NE recognition systems by using instance selectors, which can tell high-quality training sentences from noisy data. Their work inspires us to use reinforcement leaning after obtaining coarse annotated data from Gaussian-HMM.

\section{Conclusion} 
This paper presents an NE recognition model with only pre-trained word embeddings and achieves remarkable results on CoNLL 2003 English and CoNLL 2002 Spanish benchmark datasets.
The proposed approach yields, to the best of our knowledge, first fully unsupervised NE recognition work on these two benchmark datasets without any annotation data or extra knowledge base.


\bibliographystyle{acl_natbib}
\bibliography{anthology,emnlp2020}

\end{document}